\documentclass[10pt]{article}
\usepackage{latexsym,graphicx}
\newcommand{\be}{\begin{equation}}
\newcommand{\ee}{\end{equation}}
\def\n{\noindent}
\catcode `\@=11
 \catcode `\@=12
\begin{document}
\begin{center}
\large{\bf{Dark Energy Model in Anisotropic Bianchi Type-III Space-Time with Variable EoS Parameter}} \\
\vspace{10mm}
\normalsize{Anirudh Pradhan$^1$, Hassan Amirhashchi$^2$}\\
\vspace{5mm} \normalsize{$^{1}$Department of Mathematics, Hindu Post-graduate College, Zamania-232 331, 
Ghazipur, India \\
E-mail: pradhan@iucaa.ernet.in, acpradhan@yahoo.com} \\
\vspace{5mm}
\normalsize{$^{2}$Department of Physics, Mahshahr Islamic Azad University, Mahshahr, Iran \\
E-mail: hashchi@yahoo.com, h.amirhashchi@mahshahriau.ac.ir} \\
\end{center}
\vspace{10mm}
%\date{}
%\maketitle
\begin{abstract} 
A new dark energy model in anisotropic Bianchi type-III space-time with variable equation of state (EoS) parameter 
has been investigated in the present paper. To get the deterministic model, we consider that the expansion $\theta$ 
in the model is proportional to the eigen value $\sigma^{2}_{~2}$ of the shear tensor $\sigma^{j}_{~i}$. The  EoS 
parameter $\omega$ is found to be time dependent and its existing range for this model is in good agreement with the 
recent observations of SNe Ia data (Knop et al. 2003) and SNe Ia data with CMBR anisotropy and galaxy clustering 
statistics (Tegmark et al. 2004). It has been suggested that the dark energy that explains the observed accelerating 
expansion of the universe may arise due to the contribution to the vacuum energy of the EoS in a time dependent 
background. Some physical aspects of dark energy model are also discussed. 
\end{abstract}
 \smallskip
\n Keywords : Bianchi-III universe, Dark energy, Variable EoS parameter \\
\n PACS number: 98.80.Es, 95.36.+x 
%\newpage
%%%%%%%%%%%%%%%%%%%%%%%%%%%%%%%%%%%%%%%%%%%%%%%%%%%%%%%%%%%%%%%%%%%%%%%%%%%%%%%%%%%%%%%%%%%%%%%%%%%%%
%%%%%%%%%%%%%%%%%%%%%%%%%%%%%%%%%%%% SECTION 1  %%%%%%%%%%%%%%%%%%%%%%%%%%%%%%%%%%%%%%%%%%%%%%%%%%%%%
\section{Introduction}
Recent cosmological observations obtained by SNe Ia (Garnavich et al. 1998a, 1998b; Perlmutter et al. 
1997, 1998, 1999; Riess et al. 1998; Schmidt et al. 1998; Knop et al. 2003), WMAP (Bennett et al. 2003; 
Spergel et al. 2003), SDSS (Tegmark et al. 2004; Seljak et al. 2005; Adelman-McCarthy et al. 2006), 
Chandra X-ray observatory (Allen et al. 2004) indicate that the observable universe is undergoing an accelerating 
expansion. To explain the cosmic positive acceleration, mysterious dark energy has been proposed. There are 
several dark energy models which can be distinguished by, for instance, their variable equation of state (EoS)
$\omega(t) =  \frac{p}{\rho}$ (p is the fluid pressure and $\rho$ its energy density) during the evolution 
of the universe. By now, methods allowing for restoration of the quantity $\omega(t)$ from expressional data have 
been developed (Sahni and Starobinsky 2006), and an analysis of the experimental data has been conducted to determine 
this parameter as a function of cosmological time (see Sahni et al. 2008 and references therein). DE has been 
conventionally characterized by EoS parameter mentioned above which is not necessarily constant. Recently, the 
parameter $\omega(t)$ is calculated with some reasoning which reduced to some simple parametrization of the 
dependences by some authors (Huterer and Turner 2001; Weller and Albrecht 2002; Chevallier and Polarski 2001; 
Krauss et al. 2007; Usmani et al. 2008; Chen et al. 2009). The simplest DE candidate is the vacuum energy 
($\omega = - 1$), which is mathematically equivalent to the cosmological constant $\Lambda$ (Allen et al. 2004; 
Sahni and Starobinsky 2000; Sola and Stefancic 2005; Shapiro and Sola 2009). The other conventional alternatives, 
which can be described by minimally coupled scalar fields, are quintessence ($\omega > - 1$) (Ratra and Peebles 1988; 
Wetterich 1988; Liddle and Scherrer 1999), phantom energy $(\omega < - 1$) (Caldwell 2002; Caldwell et al. 2003) 
and the combination of quintessence and phantom in a unified model, namely quintom (Feng et al. 2005; Guo et al. 2005) 
as evolved and have time dependent EoS parameter. Some other limits obtained from observational results coming from 
SNe Ia data (Knop et al. 2003) and SNe Ia data collaborated with CMBR anisotropy and galaxy clustering statistics 
(Tegmark et al. 2004) are $-1.67 < \omega < -0.62$ and $-1.33 < \omega < - 0.79$ respectively. The latest results, 
obtained after a combination of cosmological datasets coming from CMB anisotropies, luminosity distances of high 
redshift type Ia supernovae and galaxy clustering, constrain the dark energy EoS to $-1.44 < \omega < -0.92$ at 
$68\%$ confidence level (Hinshaw et al. 2009; Komatsu et al. 2009). However, it is not at all obligatory to use a 
constant value of $\omega$. Due to lack of observational evidence in making a distinction between constant and variable 
$\omega$, usually the equation of state parameter is considered as a constant (Kujat et al. 2002, Bartelmann et al. 
2005) with phase wise value $-1, 0, - \frac{1}{3}$ and $ + 1$ for vacuum fluid, dust fluid, radiation and stiff 
dominated universe, respectively. But in general, $\omega$ is a function of time or redshift (Jimenez 2003; Das et al. 
2005; Ratra and Peebles 1988). For instance, quintessence models involving scalar fields give rise to time dependent 
EoS parameter $\omega$ (Turner and White 1997; Caldwell et al. 1998; Liddle and Scherrer 1999; Steinhardt et al. 1999). 
Some literature are also available on models with varying fields, such as cosmological models with variable EoS 
parameter in Kaluza-Klein metric and wormholes (Rahaman et al. 2006, 2009). In recent years various form of time 
dependent $\omega$ have been used for variable $\Lambda$ models (Mukhopadhyay et al. 2008, 2009; Usmani et al. 2008). 
Recently Ray et al. (2010), Mukhopadhyay et al. (2010), Akarsu and Kilinc (2010), Yadav (2010), Yadav \& Yadav (2010), 
Pradhan et al. (2010c) and Kumar (2010) have obtained dark energy models with variable EoS parameter in different 
contexts. \\ 

Cosmologists have proposed many candidates for dark energy to fit the current observations such as cosmological 
constant, tachyon, quintessence, phantom and so on. The major difference among these models are that they predict 
different equation of state of the dark energy and different history of the cosmos expansion. The simplest dark energy 
(DE) candidate is the cosmological constant $\Lambda$, but it needs some fine-tuning to satisfy the current value 
of the DE. Overduin and Cooperstock (1998), Sahni and Starobinsky (2000), Komatsu et al. (2009) have suggested some 
dynamic models, where $\Lambda$ varies slowly with cosmic time (t). Srivastava (2005), Jackiw (2000), Bertolami et al. 
(2004), Bento et al. (2002); Bilic et al. (2002) and Avelino et al. (2003) have considered Chaplygin gas and 
generalized Chaplygin gas as possible dark energy sources due to negative pressure. Other than these approaches, 
some authors have considered modified gravitational action by adding a function $f(R)$ (R being the Ricci scalar 
curvature) to Einstein-Hilbert Lagrangian, where $f(R)$ provides a gravitational alternative for DE causing late-time 
acceleration of the universe (Capozziello 2002; Caroll et al. 2004; Dolgov and Kawasaki 2003; Nojiri and Odintsov 
2003, 2004; Abdalaa et al. 2005; Mena et al. 2006). Recently Gupta and Pradhan (2010) have presented an entirely new 
approach as cosmological nuclear energy is a possible candidate for DE. For detail informations regarding the 
dynamics of dark energy, the readers are advised to see the reviews by Copeland et al. 2006 and Nojiri and Odintsov 
(2007). The aforementioned models offer satisfactory description of dark-energy bahaviour and its observable feathers. 
In spite of the success of these attempts, the nature of DE is one of the greatest challenge of modern cosmology.\\

Spatially homogeneous and anisotropic cosmological models play a significant role in the description of large 
scale behaviour of universe and such models have been widely studied in framework of General Relativity in search 
of a realistic picture of the universe in its early stages. Yadav et al. (2007), Pradhan et al. (2010a, 2010b) 
have recently studied homogeneous and anisotropic Bianchi type-III space-time in context of massive strings. 
Recently Yadav (2010) has obtained Bianchi type-III anisotropic DE models with constant deceleration 
parameter. In this paper, we have investigated a new anisotropic Bianchi type-III DE model with variable $\omega$ 
without assuming constant deceleration parameter. The out line of the paper is as follows: In Section $2$, the metric 
and the field equations are described. Section $3$ deals with the solution of the field equations. In Section $4$, 
some physical aspects of the derived DE model are given. Finally, conclusions are summarized in the last Section $5$.  

%%%%%%%%%%%%%%%% %%%%%%%%%%%%%%%%%%%%%%%%%%%%%%%%%%%%%%%%%%%%%%%%%%%%%%%%%%%%%%%%%%%%%%%%%%%%%%%%%%%
%%%%%%%%%%%%%%%%%%%%%%%%%%%%%%%  SECTION 2  %%%%%%%%%%%%%%%%%%%%%%%%%%%%%%%%%%%%%%%%%%%%%%%%%%%%%%%
\section{The Metric and Field  Equations}
We consider the space-time of general Bianchi-III type with the metric 
\begin{equation}
\label{eq1}
ds^{2} =  - dt^{2} + A^{2}(t) dx^{2} + B^{2}(t) e^{-2ax} dy^{2} + C^{2}(t) dz^{2},
\end{equation}
where $a$ is constant. \\ 
The simplest generalization of EoS parameter of perfect fluid may be to determine the EoS parameter separately on 
each spatial axis by preserving the diagonal form of the energy momentum tensor in a consistent way with the 
considered metric. Therefore, the energy momentum tensor of fluid is taken as
\begin{equation}
\label{eq2}
T^{j}_{i} = diag [T^{0}_{0}, T^{1}_{1}, T^{2}_{2}, T^{3}_{3}].
\end{equation}
Thus, one may parameterize it as follows,
\[
T^{j}_{i}=diag[\rho, - p_{x}, - p_{y}, - p_{z}] = diag[1, - \omega_{x}, - \omega_{y}, - \omega_{z}]\rho 
\]
\begin{equation}
\label{eq3}
= \; \ diag[1, - \omega, -(\omega + \delta), -(\omega + \gamma)]\rho.
\end{equation}
Here $\rho$ is the energy density of fluid; $p_{x}$, $p_{y}$ and $p_{z}$ are the pressures and $\omega_{x}$, 
$\omega_{y}$ and $\omega_{z}$ are the directional EoS parameters along the $x$, $y$ and $z$ axes respectively.
$\omega$ is the deviation-free EoS parameter of the fluid. We have parameterized the deviation from isotropy 
by setting $\omega_{x} = \omega$ and then introducing skewness parameter $\delta$ and $\gamma$ 
that are the deviations from $\omega$ along $y$ and $z$ axis respectively. \\\\
The Einstein's field equations (in gravitational units $c = 1$, $8\pi G = 1$) read as
\begin{equation}
\label{eq4} 
R^{j}_{i} - \frac{1}{2} R g^{j}_{i} = T^{j}_{i},
\end{equation}
where $R^{j}_{i}$ is the Ricci tensor; $R$ = $g^{ij} R_{ij}$ is the Ricci scalar. In a co-moving 
co-ordinate system, the Einstein's field equation (\ref{eq4}) with (\ref{eq3}) for the metric (\ref{eq1}) 
subsequently lead to the following system of equations:
\begin{equation}
\label{eq5}
\frac{\ddot{A}}{A} + \frac{\ddot{B}}{B} + \frac{\dot{A}\dot{B}}{AB} - \frac{a^{2}}{A^{2}} = 
- (\omega + \gamma)\rho,
\end{equation}
\begin{equation}
\label{eq6}
\frac{\ddot{B}}{B} + \frac{\ddot{C}}{C} + \frac{\dot{B}\dot{C}}{BC}  =  - \omega \rho,
\end{equation}
\begin{equation}
\label{eq7}
\frac{\ddot{A}}{A} + \frac{\ddot{C}}{C} + \frac{\dot{A}\dot{C}}{AC}  =  - (\omega + \delta)\rho,
\end{equation}
\begin{equation}
\label{eq8}
\frac{\dot{A}\dot{C}}{AC} + \frac{\dot{A}\dot{B}}{AB}  + \frac{\dot{B}\dot{C}}{BC} - \frac{a^{2}}{A^{2}} 
= \rho,
\end{equation}
\begin{equation}
\label{eq9}
\alpha\left(\frac{\dot{A}}{A} - \frac{\dot{B}}{B}\right) = 0.
\end{equation}
Here and in what follows an over dot denotes ordinary differentiation with respect to $t$. \\

The spatial volume for the model (\ref{eq1}) is given by 
\begin{equation}
\label{eq10} V^{3} = ABCe^{-ax}.
\end{equation}
We define $a = (ABCe^{-ax})^{\frac{1}{3}}$ as the average scale factor so that the Hubble's parameter is 
anisotropic models may be defined as  
\begin{equation}
\label{eq11} H = \frac{\dot{a}}{a} = \frac{1}{3}\left(\frac{\dot{A}}{A} + \frac{\dot{B}}{B} + 
\frac{\dot{C}}{C}\right).
\end{equation}
We define the generalized mean Hubble's parameter $H$ as
\begin{equation}
\label{eq12} H = \frac{1}{3}(H_{1} + H_{2} + H_{3}),
\end{equation}
where $H_{1}$ = $\frac{\dot{A}}{A}$, $H_{2}$ = $\frac{\dot{B}}{B}$ and  $H_{3} = \frac{\dot{B}}{B}$ are 
the directional Hubble's parameters in the directions of x, y and z respectively. \\

An important observational quantity is the deceleration parameter $q$, which is defined as
\begin{equation}
\label{eq13} q = - \frac{a\ddot{a}}{\dot{a}^{2}}.
\end{equation}
The scalar expansion $\theta$, components of shear $\sigma_{ij}$ and the average anisotropy parameter $Am$ 
are defined by
\begin{equation}
\label{eq14}
\theta = \frac{\dot{A}}{A} + \frac{\dot{B}}{B} + \frac{\dot{C}}{C},
\end{equation}
\begin{equation}
\label{eq15}
\sigma_{11} = \frac{A^{2}}{3}\left[\frac{2\dot{A}}{A} - \frac{\dot{B}}{B} - \frac{\dot{C}}{C}\right],
\end{equation}
\begin{equation}
\label{eq16}
\sigma_{22} = \frac{B^{2} e^{-2ax}}{3}\left[\frac{2\dot{B}}{B} - \frac{\dot{A}}{A} - \frac{\dot{C}}{C}
\right],
\end{equation}
\begin{equation}
\label{eq17}
\sigma_{33} = \frac{C^{2}}{3}\left[\frac{2\dot{C}}{C} - \frac{\dot{A}}{A} - \frac{\dot{B}}{B}\right],
\end{equation}
\begin{equation}
\label{eq18}
\sigma_{44} = 0.
\end{equation}
Therefore
\begin{equation}
\label{eq19}
\sigma^{2} = \frac{1}{3}\left[\frac{\dot{A}^{2}}{A^{2}} + \frac{\dot{B}^{2}}{B^{2}} + 
\frac{\dot{C}^{2}}{C^{2}} - \frac{\dot{A}\dot{B}}{AB} - \frac{\dot{B}\dot{C}}{BC} - 
\frac{\dot{C}\dot{A}}{CA}\right]. 
\end{equation}
\begin{equation}
\label{eq20} Am = \frac{1}{3}\sum_{i = 1}^{3}{\left(\frac{\triangle
H_{i}}{H}\right)^{2}},
\end{equation}
where $\triangle H_{i} = H_{i} - H (i = 1, 2, 3)$.
%%%%%%%%%%%%%%%%%%%%%%%%%%%%%%%%%%%%%%%%%%%%%%%%%%%%%%%%%%%%%%%%%%%%%%%%%%%%%%%%%%%%%%%%%%%%%%%%%%
%%%%%%%%%%%%%%%%%%%%%%%%%%%%%%%  SECTION 3  %%%%%%%%%%%%%%%%%%%%%%%%%%%%%%%%%%%%%%%%%%%%%%%%%%%%%%
\section{Solutions of the Field  Equations}
The field equations (\ref{eq5})-(\ref{eq9}) are a system of five equations with seven unknown parameters 
$A$, $B$, $C$, $\rho$, $\omega$, $\delta$ and $\gamma$. Two additional constraints relating these parameters 
are required to obtain explicit solutions of the system. We assume that the expansion ($\theta$) in the model 
is proportional to the eigen value 
$\sigma^{2}_{~2}$ of the shear tensor $\sigma^{j}_{~i}$. This condition leads to 
\begin{equation}
\label{eq21}
B = \ell_{1} (A C)^{m_{1}},
\end{equation}
where $\ell_{1}$ and $m_{1}$ are arbitrary constants. The motive behind assuming this condition is explained with
reference to Thorne (1967), the observations of the velocity-red-shift relation for extragalactic 
sources suggest that Hubble expansion of the universe is isotropic today within $\approx 30$ per cent 
(Kantowski and Sachs 1966; Kristian and Sachs 1966). To put more precisely, red-shift studies place the limit
$$
\frac{\sigma}{H} \leq 0.3,
$$
on the ratio of shear $\sigma$ to Hubble constant $H$ in the neighbourhood of our Galaxy today. Collins 
et al. (1980) have pointed out that for spatially homogeneous metric, the normal congruence to the 
homogeneous expansion satisfies that the condition $\frac{\sigma}{\theta}$ is constant. \\\\ 
Equations (\ref{eq9}) leads to
\begin{equation}
\label{eq22}
A = m B,
\end{equation}
where $m$ is a positive integrating constant. We have revisited the solution recently obtained by Pradhan 
et al. (2010a).\\\\
Using the value of Eq. (\ref{eq22}) in (\ref{eq7}) and subtract the result from equation (\ref{eq6}), we obtain 
the skewness parameter on y-axis is null i.e. $\delta = 0$. In this case Eqs. (\ref{eq6}) and (\ref{eq7}) 
are reduced to
\begin{equation}
\label{eq23}
\frac{\ddot{B}}{B} - \frac{\ddot{A}}{A} + \frac{\dot{B}\dot{C}}{BC} - \frac{\dot{A}\dot{C}}{AC} = 0.
\end{equation}
Using (\ref{eq22}) in (\ref{eq23}), we obtain
\begin{equation}
\label{eq24}
(1 - m)\left(\frac{\ddot{B}}{B} + \frac{\dot{B}\dot{C}}{BC}\right) = 0.
\end{equation}
As $m \ne 0$, Eq. (\ref{eq24}) gives
\begin{equation}
\label{eq25}
\left(\frac{\ddot{B}}{B} + \frac{\dot{B}\dot{C}}{BC}\right) = 0,
\end{equation}   
which on integration reduces to
\begin{equation}
\label{eq26}
\dot{B}C = k_{1},
\end{equation}
where $k_{1}$ is an integrating constant. \\
From Eqs. (\ref{eq21}) and (\ref{eq22}), we obtain
\begin{equation}
\label{eq27}
B = \ell_{2} C^{\ell},
\end{equation}
where $\ell_{2} = \ell_{1}^{\frac{1}{1 - m_{1}}} m^{\ell}$, $\ell = \frac{m_{1}}{1 - m_{1}}$. Using (\ref{eq27})
in (\ref{eq26}), we get
\begin{equation}
\label{eq28}
C^{\ell} \dot{C} = \frac{k_{1}}{\ell \ell_{2}},
\end{equation}
which on integration gives
\begin{equation}
\label{eq29}
C = (\ell + 1)^{\frac{1}{\ell + 1}}\left[\frac{k_{1}}{\ell \ell_{2}}t + k_{2}\right]^{\frac{1}{\ell + 1}},
\end{equation}
where $k_{2}$ is an integrating constant. Using (\ref{eq29}) in (\ref{eq27}) and (\ref{eq22}) we obtain
\begin{equation}
\label{eq30}
B = \ell_{2} (\ell + 1)^{\frac{\ell}{\ell + 1}}\left[\frac{k_{1}}{\ell \ell_{2}}t + k_{2}\right]^{\frac{\ell}
{\ell + 1}},
\end{equation}
and
\begin{equation}
\label{eq31}
A = m\ell_{2}(\ell + 1)^{\frac{\ell}{\ell + 1}}\left[\frac{k_{1}}{\ell \ell_{2}}t + k_{2}\right]^{\frac{\ell}
{\ell + 1}},
\end{equation}
respectively. \\
Hence the metric (\ref{eq1}) reduces to the form
\[
ds^{2} = - dt^{2} + \Biggl[m\ell_{2}(\ell + 1)^{\frac{\ell}{\ell + 1}}\left(\frac{k_{1}}{\ell \ell_{2}}t + 
k_{2}\right)^{\frac{\ell}{\ell + 1}}\Biggr]^{2}dx^{2} + 
\]
\begin{equation}
\label{eq32}
\Biggl[\ell_{2} (\ell + 1)^{\frac{\ell}{\ell + 1}}e^{-ax}\left(\frac{k_{1}}{\ell \ell_{2}}t + k_{2}
\right)^{\frac{\ell}{\ell + 1}}\Biggr]^{2}dy^{2} + \Biggl[(\ell + 1)^{\frac{1}{\ell + 1}}
\left(\frac{k_{1}}{\ell \ell_{2}}t + k_{2}\right)^{\frac{1}{\ell + 1}}\Biggr]^{2}dz^{2}.
\end{equation}
Using the suitable transformation
$$m\ell_{2}(\ell + 1)^{\frac{\ell}{\ell + 1}} x = X,$$
$$\ell_{2}(\ell + 1)^{\frac{\ell}{\ell + 1}} y = Y,$$
$$ (\ell + 1)^{\frac{1}{\ell + 1}}z = Z,$$
\begin{equation}
\label{eq33}
\frac{k_{1}}{\ell \ell_{2}} t = T, 
\end{equation}
the metric (\ref{eq32}) is reduced to  
\begin{equation}
\label{eq34}
ds^{2} = -\beta^{2}dT^{2} + T^{2L}dX^{2} + T^{2L}e^{-\frac{2a}{N}X}dY^{2} + T^{\frac{2L}{\ell}}dZ^{2},
\end{equation}
where 
$$\beta = \frac{\ell\ell_{2}}{k_{1}},$$
$$M = (\ell + 1)^{\frac{1}{\ell + 1}},$$
$$N = m\ell_{2}M,$$
\begin{equation}
\label{eq35}
L = \frac{\ell}{\ell + 1}.
\end{equation}
%%%%%%%%%%%%%%%%%%%%%%%%%%%%%%%%%%%%%%%%%%%%%%%%%%%%%%%%%%%%%%%%%%%%%%%%%%%%%%%%%%%%%%%%%%%%%%%%%%
%%%%%%%%%%%%%%%%%%%%%%%%%%%%%%%  SECTION 4  %%%%%%%%%%%%%%%%%%%%%%%%%%%%%%%%%%%%%%%%%%%%%%%%%%%%%%
\section{Some Physical and Geometric Properties of the Model}
The expressions for the scalar of expansion $\theta$, magnitude of shear $\sigma^{2}$, the average 
anisotropy parameter $A_{m}$, deceleration parameter $q$ and proper volume $V$ for the DE model 
(\ref{eq34}) are given by
\begin{equation}
\label{eq36} \theta = \frac{(2\ell + 1)L}{\ell \beta T},
\end{equation}
\begin{equation}
\label{eq37} \sigma^{2} = \frac{1}{3}\left(\frac{(\ell - 1)L}{\ell \beta T}\right)^{2},
\end{equation}
\begin{equation}
\label{eq38} A_{m} =  2\left(\frac{\ell - 1}{2\ell + 1}\right)^{2},
\end{equation}
\begin{equation}
\label{eq39} q = -\frac{\ell \beta}{(2\ell + 1)},
\end{equation}
\begin{equation}
\label{eq40} V =  \frac{N^{2}}{m}M^{\frac{1}{\ell}}T^{\frac{L(2\ell + 1)}{\ell}}.
\end{equation}
The rate of expansion $H_{i}$ in the direction of x, y and z are
given by
\begin{equation}
\label{eq41} H_{1} = H_{2} = \frac{L}{\beta T},
\end{equation}
\begin{equation}
\label{eq42} H_{3} = \frac{L}{\ell \beta T}.
\end{equation}
Hence the average generalized Hubble's parameter is given by
\begin{equation}
\label{eq43} H =  \frac{L(2\ell + 1)}{3\ell \beta T}.
\end{equation}
%%%%%%%%%%%%%%%%%%% Figure 1 %%%%%%%%%%%%
\begin{figure}[ht]
\centering
\includegraphics[width=10cm,height=10cm,angle=0]{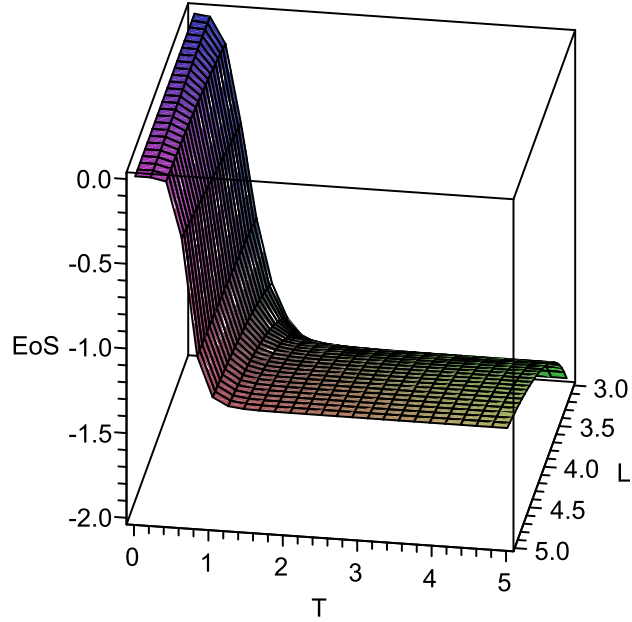} \\
\caption{The plot of EoS parameter $\omega$ versus $T$ and $L$}
\end{figure}
%%%%%%%%%%%%%%%%%%%%%%%%%%%%%%%%%% %%%%%%%%
%%%%%%%%%%%%%%%%%%% Figure 2 %%%%%%%%%%%%
\begin{figure}[ht]
\centering
\includegraphics[width=10cm,height=10cm,angle=0]{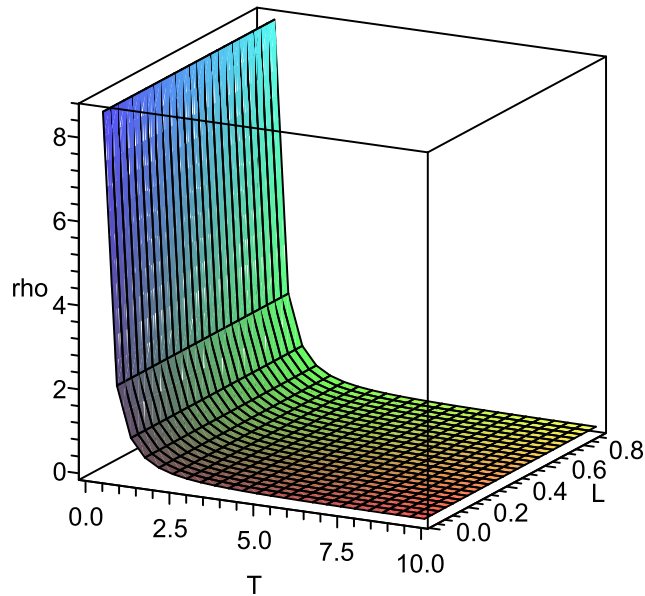} \\
\caption{The plot of energy density $\rho$ versus $T$ and L}
\end{figure}
%%%%%%%%%%%%%%%%%%%%%%%%%%%%%%%%%% %%%%%%%%
Using equations (\ref{eq29}) - (\ref{eq31}) in (\ref{eq8}), the energy density of the fluid is 
obtained as
\begin{equation}
\label{eq44}
\rho = \frac{L^{2}(\ell + 2)}{\ell \beta^{2}T^{2}} - \frac{a^{2}}{N^{2}T^{2L}}.
\end{equation}
Now, by using equations (\ref{eq29}), (\ref{eq30}) and (\ref{eq44}) in (\ref{eq6}), the equation of 
state parameter $\omega$ is obtained as
\begin{equation}
\label{eq45}  \omega = \frac{L (\ell^{2} + \ell + 1) - \ell(\ell + 1)}{\frac{\ell^{2} a^{2} 
\beta^{2}}{N^{2} L T^{2(L - 1)}} - L \ell (\ell + 2)}.
\end{equation}
Using equations (\ref{eq30}), (\ref{eq31}), (\ref{eq44}) and (\ref{eq45}) in (\ref{eq5}), the skewness 
parameter $\gamma$ (i.e. deviation from $\omega$ along z-axis) is derived as
\begin{equation}
\label{eq46} \gamma = \frac{\frac{a^{2}}{N^{2}T^{2L}} + \frac{L}{\ell^{2}\beta^{2}T^{2}}[\ell(\ell - 1) 
- L(2\ell^{2} - \ell -1)]}{\frac{L^{2}(\ell + 2)}{\ell \beta^{2}T^{2}} - \frac{a^{2}}{N^{2}T^{2L}}}.
\end{equation}
From equation (\ref{eq45}), it is observed that the equation of state parameter $\omega$ is time
dependent, it can be function of redshift $z$ or scale factor $a$ as well. The redshift
dependence of $\omega$ can be linear like 
\begin{equation}
\label{eq47}
\omega(z) = \omega_{0} + \omega^{'} z,
\end{equation}
with $\omega^{'}$ = $(\frac{d\omega}{dz})_{z} = 0$ (see Refs. Huterer and Turner 2001; Weller and Albrecht 
2002) or nonlinear as 
\begin{equation}
\label{eq48}
\omega(z) = \omega_{0} + \frac{\omega_{1}z}{1 + z},
\end{equation}
(Polarski and Chavellier 2001; Linder 2003). So, as for as the scale factor dependence of $\omega$ is concern, the 
parametrization is given by
\begin{equation}
\label{eq49}
\omega(a) = \omega_{0} + \omega_{a}(1 - a),
\end{equation}
where $\omega_{0}$ is the present value ($a = 1$) and $\omega_{a}$ is the measure of the time variation $\omega^{'}$ 
(Linder 2008). \\\\
The SNe Ia data suggests that $-1.67 < \omega < -0.62$ (Knop et al. 2003) while the limit imposed on $\omega$ by 
a combination of SNe Ia data with CMB anisotropy and galaxy clustering statistics is $-1.33 < \omega < -0.79$ 
(Tegmark et al. 2004). So, if the present work is compared with experimental results mentioned above, then one can 
conclude that the limit of $\omega$ provided by equation (\ref{eq45}) may accommodated with the acceptable range 
of EoS parameter. Also it is observed that either for $T = 0$ or for $m_{1} = 0$, the $\omega$ vanishes and our 
model represents a dusty universe.\\\\
For the value of $\omega$ to be in consistent with observation (Knop et al. 2003), we have the following 
general condition
\begin{equation}
\label{eq50}
T_{1} < T < T_{2},
\end{equation}
where
\begin{equation}
\label{eq51}
T_{1} = \left[\frac{0.79 \ell a \beta}{N\sqrt{L\{\ell(\ell + 1) - L(0.38 \ell^{2} - 0.24 \ell + 1)\}}}\right]^{\frac{1}
{(1 - L)}},
\end{equation}
and
\begin{equation}
\label{eq52}
T_{2} = \left[\frac{1.3 \ell a \beta}{N\sqrt{L\{\ell(\ell + 1) - L(0.67 \ell^{2} - 2.34 \ell + 1)\}}}\right]^{\frac{1}
{(1 - L)}}.
\end{equation}
For this constrain, we obtain  $-1.67 < \omega < -0.62$, which is in good agreement with the limit obtained from 
observational results coming from SNe Ia data (Knop et al. 2003). For a special case for which $\ell = 1$, 
$a = 0.5$, $\beta = 2$, $L = 0.5$, $N = 1$, where $0.899700 < T < 1.153226$, we obtain the same limit 
 $-1.67 < \omega < -0.62$. \\\\
From Eq. (\ref{eq45}), we have observed that, at cosmic time
\begin{equation}
\label{eq53}
T = \left[\frac{\ell a \beta}{N\sqrt{L\{L\ell + \ell(\ell + 1) - 1\}}}\right]^{\frac{1}
{(L - 1)}},
\end{equation}
$\omega = - 1$ (i.e. cosmological constant dominated universe) and when 
\begin{equation}
\label{eq54}
T < \left[\frac{\ell a \beta}{N\sqrt{L\{L\ell + \ell(\ell + 1) - 1\}}}\right]^{\frac{1}
{(L - 1)}},
\end{equation}
$\omega > -1$ (i.e. quintessence) and when
\begin{equation}
\label{eq55}
T > \left[\frac{\ell a \beta}{N\sqrt{L\{L\ell + \ell(\ell + 1) - 1\}}}\right]^{\frac{1}
{(L - 1)}},
\end{equation}
$\omega < -1$ (i.e. super quintessence or phantom fluid dominated universe) (Caldwell 2002). \\\\
The variation of EoS parameter $\omega$ with cosmic time $T$ is clearly shown in {\bf 
Figures 1}, as a representative case with appropriate choice of constants of integration and other physical 
parameters using reasonably well known situations. From {\bf Figure 1}, we conclude that in early stage of 
evolution of the universe, the EoS parameter $\omega$ was very small but positive (i.e, the universe was 
matter dominated) and at late time it is evolving with negative value (i.e. at the present time). The earlier
real matter later on converted to the dark energy dominated phase of the universe.\\\\
From Eq. (\ref{eq44}), we note that $\rho(t)$ is a decreasing function of time and $\rho > 0$ for all times. 
This behaviour is clearly depicted in {\bf Figures $2$} as a representative case with appropriate choice of 
constants of integration and other physical parameters using reasonably well known situations. \\\\
In absence of any curvature, matter energy density ($\Omega_{m}$) and dark energy ($\Omega_{\Lambda}$) are 
related by the equation
\begin{equation}
\label{eq56} \Omega_{m} + \Omega_{\Lambda} = 1,
\end{equation}
where $\Omega_{m} = \frac{\rho}{3H^{2}}$ and $\Omega_{\Lambda} = \frac{\Lambda}{3H^{2}}$.
Thus, equation (\ref{eq56}) reduces to
\begin{equation}
\label{eq57} \frac{\rho}{3H^{2}} + \frac{\Lambda}{3 H^{2}} = 1.
\end{equation}
Using equations (\ref{eq43}) and (\ref{eq44}), in equation (\ref{eq57}), the cosmological constant is
obtained as
\begin{equation}
\label{eq58} \Lambda = \frac{{L}^{2}(2\ell + 1)^{2}}{3\ell^{2} \beta^{2} T^{2}} - 
{\frac{L^{2}(\ell + 2)}{\ell \beta^{2} T^{2}}} + {\frac{a^{2}}{N^{2}T^{2 L}}}.
\end{equation}
%%%%%%%%%%%%%%%%%%% Figure 3 %%%%%%%%%%%%
\begin{figure}[ht]
\centering
\includegraphics[width=10cm,height=10cm,angle=0]{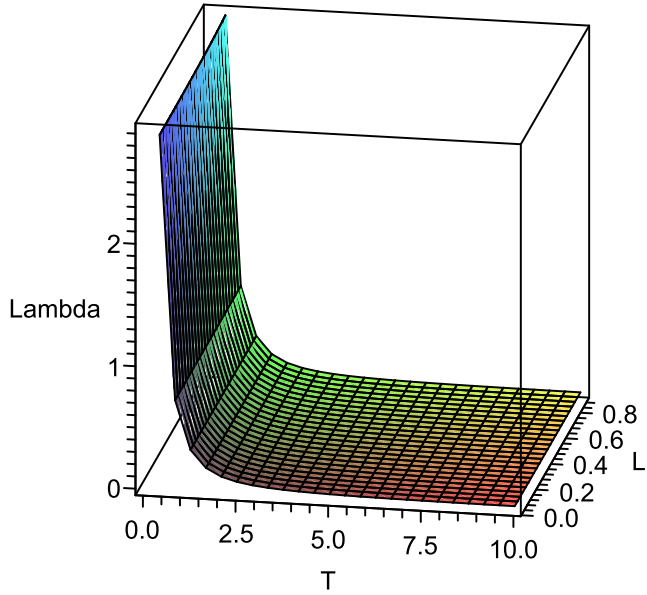} \\
\caption{The plot of cosmological term $\Lambda$ versus $T$ and L}
\end{figure}
%%%%%%%%%%%%%%%%%%%%%%%%%%%%%%%%%% %%%%%%%%
In recent time the $\Lambda$-term has interested theoreticians and observers for various reasons. The 
nontrivial role of the vacuum in the early universe generate a $\Lambda$-term that leads to inflationary 
phase. Observationally, this term provides an additional parameter to accommodate conflicting data on the 
values of the Hubble constant, the deceleration parameter, the density parameter and the age of the universe 
(for example, see the references Gunn and Tinsley 1975; Wampler and Burke 1988). The behaviour of the 
universe in this model will be determined by the cosmological term $\Lambda$ ; this term has the same effect 
as a uniform mass density $\rho_{eff} = - \Lambda$, which is constant in time. A positive 
value of $\Lambda$ corresponds to a negative effective mass density (repulsion). Hence, we expect that in 
the universe with a positive value of $\Lambda$, the expansion will tend to accelerate; whereas in the 
universe with negative value of $\Lambda$, the expansion will slow down, stop and reverse. From Eq. 
(\ref{eq58}), we see that the cosmological term $\Lambda$ is a decreasing function of time and it approaches 
a small positive value at late time. From {\bf Figure 3}, we note this behaviour of cosmological term $\Lambda$ 
in the model. Recent cosmological observations  suggest the existence of a positive cosmological constant 
$\Lambda$ with the magnitude $\Lambda(G\hbar/c^{3})\approx 10^{-123}$. These observations on magnitude and 
red-shift of type Ia supernova suggest that our universe may be an accelerating one with induced cosmological 
density through the cosmological $\Lambda$-term. It is remarkable to mention here that the dark energy that explains 
the observed accelerating expansion of the universe may arise due to the contribution to the vacuum energy of the 
EoS in a time dependent background. Thus, our DE model is consistent with the results of recent observations. \\\\  
From the above results, it can be seen that the spatial volume is zero at $T = 0$ and it increases 
with the increase of T. This shows that the universe starts evolving with zero volume at $T = 0$ and 
expands with cosmic time T. From equations (\ref{eq41}) and (\ref{eq42}), we observe that all the three 
directional Hubble parameters are zero at $T \to \infty$. In derived model, the energy density tend 
to infinity at $T = 0$. The model has the point-type singularity at $T = 0$ (MacCallum 1971). The shear 
scalar diverses at $T = 0$. As $T \to \infty$, the scale factors $A(t)$, $B(t)$ and $C(t)$ tend to 
infinity. The energy density becomes zero as $T \to \infty$. The expansion scalar and shear scalar all 
tend to zero as $T \to \infty$. The mean anisotropy parameter are uniform throughout whole expansion 
of the universe when $\ell \ne -\frac{1}{2}$ but for $\ell = - \frac{1}{2}$ it tends to infinity. This shows 
that the universe is expanding with the increase of cosmic time but the rate of expansion and shear 
scalar decrease to zero and tend to isotropic. At the initial stage of expansion, when $\rho$ is large, 
the Hubble parameter is also large and with the expansion of the universe $H$, $\theta$ decrease as 
does $\rho$. Since $\frac{\sigma^{2}}{\theta^{2}} = $ constant provided $\ell \ne - \frac{1}{2}$, the model 
does not approach isotropy at any time. The cosmological evolution of Bianchi type-III space-time is 
expansionary, with all the three scale factors monotonically increasing function of time. The dynamics 
of the mean anisotropy parameter depends on the value of $\ell$. \\\\
From (\ref{eq39}) we observe that 
\[
 (i) ~ ~ ~ ~ ~ ~ for ~ ~ ~ \ell < - \frac{1}{2}, ~ ~ q > 0  
\]
i.e., the model is decelerating and 
\[
(ii) ~ ~ ~ ~ ~ ~ for ~ ~ ~ \ell > - \frac{1}{2}, ~ ~ q < 0  
\]
i.e., the model is accelerating. Thus this case implies an accelerating model of the universe. Recent 
observations of type Ia supernovae (see Perlmutter et al., 1999; Riess et al., 1998 and references therein) 
reveal that the present universe is in accelerating phase and deceleration parameter lies somewhere in 
the range $-1 < q \leq 0$. It follows that our DE model of the universe is consistent with the recent 
observations.
%%%%%%%%%%%%%%%%%%%%%%%%%%%%%%%%%%%%%%%%%%%%%%%%%%%%%%%%%%%%%%%%%%%%%%%%%%%%%%%%%%%%%%%%%%%%%%%%%%
%%%%%%%%%%%%%%%%%%%%%%%%%%%%%%%  SECTION 5  %%%%%%%%%%%%%%%%%%%%%%%%%%%%%%%%%%%%%%%%%%%%%%%%%%%%%%
\section{Concluding Remarks}
An anisotropic Bianchi type-III DE model with variable EoS parameter $\omega$ has been investigated 
which is new and different from the other author's solutions. In the derived model, $\omega$ is obtained as 
time varying which is consistent with recent observations (Knop et al. 2003; Tegmark et al. 2004). It is observed 
that, in early stage, the equation of state parameter $\omega$ is positive i.e. the universe was matter dominated 
in early stage but in late time, the universe is evolving with negative values i.e the present epoch (see, 
{\bf Figure 1}). Our DE model is in accelerating phase which is consistent with the recent observations. Thus the 
model (\ref{eq34}) represents a realistic model. \\

In the derived DE model of the universe, the cosmological term is a decreasing function of time and it approaches 
a small positive value at late time (i.e., the present epoch). The values of cosmological ``constant'' 
for the model is found to be small and positive, which is supported by the recent observations (Garnavich 
et al. 1998a, 1998b; Perlmutter et al. 1997, 1998, 1999; Riess et al. 1998, 2000, 2004; Schmidt et al. 1998). \\

The DE model is based on exact solution of Einstein's field equations for the anisotropic Bianchi-III space-time 
filled with perfect fluid. To my knowledge, the literature has hardly witnessed this sort of exact solution for 
anisotropic Bianchi-III space-time. So the derived DE model adds one more feather to the literature. \\

The DE model presents the dynamics of EoS parameter $\omega$ provided by Eq. (\ref{eq45}) may accommodated with 
the acceptable range  $-1.67 < \omega < -0.62$ of SNe Ia data (Knop et al. 2003). It is already observed and shown 
in previous section that for different cosmic times, we obtain cosmological constant dominated universe, quintessence 
and phantom fluid dominated universe (Caldwell 2002), representing the different phases of the universe through out 
the evolving process. Therefore, we can not rule out the possibility of anisotropic nature of DE at least in 
the framework of Bianchi-III space-time. \\

Though there are many suspects (candidates) such as cosmological constant, vacuum energy, scalar field, brane world, 
cosmological nuclear-energy, etc. as reported in the vast literature for DE, the proposed model in this paper at 
least presents a new candidate (EoS parameter) as a possible suspect for the DE.   

\section*{Acknowledgments} 
The first author (A. Pradhan) would like to thank the Inter-University Centre for Astronomy and Astrophysic (IUCAA), 
Pune, India for providing facility under associateship programme where part of this work was carried out. The authors 
also thank the anonymous referee for constructive suggestions.

\end{document}